\documentclass[runningheads]{llncs}
\usepackage[T1]{fontenc}
\usepackage{amsmath}
\usepackage{hyperref}
\usepackage{amsmath, amssymb}
\usepackage{graphicx}
\begin{document}
\title{Can Differentially Private Fine-tuning LLMs Protect Against  Privacy Attacks?}
%

% If the paper title is too long for the running head, you can set
% an abbreviated paper title here
%
\author{Hao Du\inst{1} \and Shang Liu\inst{2} \and Yang Cao\inst{3}}
\authorrunning{H. Du \and S. Liu et al.}
\authorrunning{Hao Du et al.}
% First names are abbreviated in the running head.
% If there are more than two authors, 'et al.' is used.
%
\institute{Hokkaido University \and China University of Mining and Technology \and Institute of Science Tokyo}
\maketitle              % typeset the header of the contribution
\begin{abstract}
Fine-tuning large language models (LLMs) has become an essential strategy for adapting them to specialized tasks; however, this process introduces significant privacy challenges, as sensitive training data may be inadvertently memorized and exposed. 
Although differential privacy (DP) offers strong theoretical guarantees against such leakage, its empirical privacy effectiveness on LLMs remains unclear, especially under different fine-tuning methods. 
In this paper, we systematically investigate the impact of DP across fine-tuning methods and privacy budgets, using both data extraction and membership inference attacks to assess empirical privacy risks.
Our main findings are as follows:
(1) Differential privacy reduces model utility, but its impact varies significantly across different fine-tuning methods.
(2) Without DP, the privacy risks of models fine-tuned with different approaches differ considerably.
(3) When DP is applied, even a relatively high privacy budget can substantially lower privacy risk.
(4) The privacy-utility trade-off under DP training differs greatly among fine-tuning methods, with some methods being unsuitable for DP due to severe utility degradation.
Our results provide practical guidance for privacy-conscious deployment of LLMs and pave the way for future research on optimizing the privacy-utility trade-off in fine-tuning methodologies.

\end{abstract}
\section{Introduction}
In recent years, large language models (LLMs), including GPT-4, LLaMA, and PaLM, have significantly advanced natural language processing (NLP), enabling diverse applications such as content generation, machine translation, question answering, and code completion. 
These models achieve remarkable proficiency in understanding and generating human-like text by training on extensive datasets, typically drawn from publicly available internet resources. 
Despite their impressive generalization abilities, LLMs often exhibit suboptimal performance when directly applied to specialized domains or specific tasks without additional adaptation or fine-tuning.

Fine-tuning has become a crucial approach for adapting large language models (LLMs) to specialized downstream tasks, enabling these models to achieve state-of-the-art performance across various fields. 
However, as modern LLMs continue to grow in scale, fully fine-tuning their large number of parameters can lead to excessive computational resource consumption. 
To address this issue, parameter-efficient fine-tuning (PEFT) methods \cite{han2024parameterefficient,liu2022fewshot} have been introduced, significantly reducing the number of adjustable parameters and consequently lowering resource requirements. 
Nevertheless, fine-tuning often involves sensitive data, raising considerable privacy concerns, such as vulnerability to data extraction and membership inference attacks, which can compromise the confidentiality of information used during the fine-tuning process.
% \vspace{-0.3em}

In this context, differentially private optimization algorithms have emerged as a widely adopted approach to mitigate privacy risks during the fine-tuning of large language models (LLMs). 
These algorithms integrate differential privacy (DP) principles by introducing carefully calibrated noise into gradient updates during training, effectively restricting the influence of individual data points on the resulting model parameters. 
A prominent example is differentially private stochastic gradient descent (DP-SGD)~\cite{abadi2016deep}, which modifies traditional stochastic optimization by clipping per-sample gradients and adding calibrated noise. 
Consequently, this method ensures that the contribution of any individual data point remains indistinguishable within the overall model behavior. 
DP-SGD thus offers strong privacy guarantees, enabling models to leverage sensitive datasets without compromising individual data privacy.

Although differential privacy offers strong theoretical privacy guarantees, its empirical privacy effectiveness on LLMs remains unclear, especially under different fine-tuning methods.
Existing studies \cite{fu2023practical,lukas2023analyzing,marchyok2025evaluating,mireshghallah-etal-2022-empirical,panda2025privacy} have not thoroughly examined how DP influences empirical privacy risks or its effectiveness in mitigating a range of privacy attacks.
Lukas et al. \cite{lukas2023analyzing} investigated the leakage of personally identifiable information (PII) in large language models and assessed DP's impact on privacy risks. However, their study focused solely on full fine-tuning and considered only a single privacy budget scenario, without exploring PEFT approaches.
Similarly, Fu et al. \cite{fu2023practical} evaluated the defense provided by DP training against membership inference attacks but limited their experiments exclusively to the LoRA method.
Panda et al. \cite{panda2025privacy} examined privacy auditing methods in large language models; however, their evaluation was constrained to a single privacy budget and lacked diversity in fine-tuning methods.
Although some analyses have addressed model privacy risks across different fine-tuning strategies, these studies exhibit various shortcomings. Mireshghallah et al. \cite{mireshghallah-etal-2022-empirical} conducted a comprehensive analysis of memorization in autoregressive models; however, due to the age of their work, the membership inference attack techniques and fine-tuning methods they employed are now outdated.
Additionally, one paper \cite{marchyok2025evaluating} focused specifically on the privacy risks of different PEFT methods. However, their experiments primarily relied on the exposure metric from data extraction attacks to evaluate model memorization, which does not provide a complete picture.
Notably, the experimental results for LoRA and prefix-tuning reported in these studies differ from our findings, as detailed in \textit{Remark \ref{remark:privacy budget}}.

To address the gap between theoretical guarantees and real-world privacy attacks, this paper systematically investigates the impact of differential privacy (DP) on fine-tuning large language models (LLMs), with a particular focus on both full and parameter-efficient fine-tuning (PEFT) methods. 
We evaluate how different privacy budgets affect model utility and empirical privacy risk using two representative attack techniques: data extraction and membership inference. 
Our study spans multiple fine-tuning strategies and model sizes to ensure broad applicability.
Key contributions of this work include: 
\begin{itemize}
\item[(1)] A comprehensive comparison of the privacy-utility trade-off across fine-tuning methods under DP constraints.
\item[(2)] Empirical evidence showing that full fine-tuning and LoRA offer favorable trade-offs, while prefix-tuning suffers from severe utility degradation.
\item[(3)] Practical insights for selecting fine-tuning strategies in privacy-sensitive applications. These findings serve as a foundation for optimizing private fine-tuning in future LLM deployments.
\end{itemize}

\section{Related Work}
\subsection{Parameter-efficient Fine-tuning (PEFT)}
Fine-tuning is a fundamental process for adapting pre-trained models to downstream tasks, enabling them to leverage the vast knowledge captured during pre-training. 
While traditional full fine-tuning updates all parameters of a model, achieving high task-specific performance, it is computationally expensive and prone to overfitting, especially with limited data. 
To overcome these challenges, many Parameter-Efficient Fine-Tuning (PEFT) methods \cite{ben-zaken-etal-2022-bitfit,hu2022lora,lester-etal-2021-power,li-liang-2021-prefix,liu2024gpt,xu2023parameter} have been proposed, offering more efficient alternatives. 
These methods vary in their approach, striking different balances between parameter efficiency, adaptability, performance, and speed, expand the range of available fine-tuning methods.

Full Fine-tuning~\cite{devlin2018bert} (FFT) updates all parameters of a model, providing maximum flexibility and allows the model to fully adapt to the task. However, it is computationally intensive and requires substantial data to avoid overfitting, making it less practical for resource-constrained scenarios. 
Prefix-tuning~\cite{li-liang-2021-prefix} adds trainable prefix vectors at every Transformer layer, interacting with input through attention mechanisms. Compared to Prompt-tuning, it allows dynamic adjustment of intermediate representations, improving adaptability for complex tasks, but slightly increases computational cost due to its per-layer modifications.
LoRA~\cite{hu2022lora} uses low-rank decomposition of weight matrices and fine-tunes only the low-rank components. It maintains a effective balance between parameter efficiency and task performance, making it highly applicable in large-scale models and complex tasks.
P-tuning~\cite{liu2024gpt} addresses the limitation of traditional prompt-based learning which is unstable by introducing continuous prompt embeddings that are learned during training. 
These embeddings are concatenated with discrete prompts and input tokens, allowing the model to adapt more effectively to specific tasks.
\subsection{Differential Privacy (DP)}
Differential privacy (DP)~\cite{dwork2006differential} was originally developed to provide strong privacy guarantees in statistical databases by ensuring that query results do not reveal sensitive information about any individual record.
DP-SGD (Differentially Private Stochastic Gradient Descent)~\cite{abadi2016deep} incorporates this concept into training by clipping gradients and injecting noise, limiting the influence of any single data point.
After DP-SGD, many DP training methods aim to reduce the computational and memory overhead of per-example gradient clipping, enhance efficiency, and improve the trade-off between privacy and utility. 
Ghost Clipping~\cite{8970912} approximates the per-sample clipping without explicitly calculating each gradient, thereby reducing the computational and memory overhead.

\textbf{Book Keeping}~\cite{10.5555/3618408.3618538} works by recording and reusing the output gradients computed during the initial backward pass. 
This avoids the need for a second back-propagation (required by earlier methods like GhostClip) to compute per-sample gradient norms. 
By "book-keeping" the output gradients, the method significantly reduces both time and memory overhead, bringing the efficiency of differentially private training closer to that of standard (non-DP) training.
In our work, we employed the Book Keeping approach for DP training to ensure differential privacy. Specifically, we utilized the open-source fastDP library to implement this method. 

\subsection{Membership inference attack} 
Membership inference attacks (MIAs) are designed to determine whether specific data points were included in a model's training set, thereby posing considerable risks to sensitive information such as clinical records and user preference datasets. 
Recent research has investigated the underlying mechanisms, vulnerabilities, and effects of MIAs on large language models (LLMs), illuminating various attack strategies and factors that affect model susceptibility. 
Mireshghallah et al.~\cite{mireshghallah2022empirical} proposed an MIA based on Likelihood Ratios, using the original model as a reference. 
Their work also conducted a preliminary investigation into how different fine-tuning methods affect the effectiveness of the attacks.
Jagannatha et al.~\cite{jagannatha2021membership} conducted a black-box MIA on clinical language models, employing a Threshold-Based Attack to identify which samples were included in fine-tuning across models of varying sizes. 

\textbf{SPV-MIA}~\cite{fu2023practical} further increased the AUC for membership inference attacks on large language models to over 90\%. 
Their method eliminates the need for attackers to access an external reference dataset. 
Instead, the adversary prompts the target LLM to generate texts, which form a reference dataset used to fine-tune a reference model.
By comparing the target model’s sampling probability with that of the reference model, the method calibrates the inherent biases.
Specifically, in the probabilistic variation assessment, it generates slight, symmetrical paraphrases of a text to approximate the local variation (similar to a second derivative) of the probability function.
This variation signal, after calibration, serves as a robust indicator to determine if the text was part of the training set.
In our work, we chose SPV-MIA as the membership inference attack method to evaluate the model's privacy risk.

\subsection{Data extraction attack}
Data extraction and reconstruction attacks reveal a critical vulnerability in large language models (LLMs), enabling adversaries to retrieve sensitive data, including personally identifiable information (PII), from both the models' outputs and internal representations. 
These vulnerabilities underscore the importance of assessing security risks when deploying LLMs in sectors that handle sensitive or proprietary information, such as healthcare, finance, and customer service. 
Recent research has explored a variety of data extraction techniques, highlighting the increasing sophistication and success of these attacks.
Early research on data extraction primarily focused on retrieving pre-training data. Carlini et al.~\cite{48112} attempted to extract sensitive information that was inadvertently memorized during language model training. They also introduced the \textit{exposure} metric to quantify the extent to which a model memorizes specific data, significantly influencing subsequent research.
Lukas et al.~\cite{lukas2023analyzing} focused on extracting personal identifiable information (PII) from models under varying conditions, proposing more advanced extraction and reconstruction attacks.
In our work, we employed a basic prompt attack and used the exposure metric to quantify risk, which will be detailed in the next section.

\section{Methodology}\label{section3}
In this paper, we evaluate the privacy risks of language models by implementing two types of privacy attacks: a data extraction attack and a membership inference attack (MIA). For the data extraction attack, we employ a prompt attack to attempt to extract a canary embedded in the fine-tuning dataset. Regarding the MIA, we adopt SPV-MIA, the current state-of-the-art method specifically designed for language models. To protect the fine-tuning dataset, we utilize 
\subsection{Data Extraction Attack}
\subsubsection{Adversary's Capability.} 
Our method simulates a typical black-box adversary. 
In this setting, the adversary can submit crafted prompts to the model and observe its outputs. 
Additionally, we assume that the adversary has partial knowledge of the fine-tuning dataset. 
This may include access to some of the training samples or general information about the data distribution. 
This combination of query access and auxiliary data constitutes the adversarial capability in our experiments.
\subsubsection{Threat Model.}
We designed a simple prompt-based method to simulate a data extraction scenario for evaluating model privacy risk. 
First, we selected an open-source dataset and inserted repeated canary samples, which is equivalent to 0.25\% of the total training samples, into its training set. 
The 0.25\% proportion was chosen because it does not overly distort the dataset distribution while being sufficient for the model to memorize the canary, as evidenced by the fact that, in the absence of differential privacy, full fine-tuning can fully output the canary sample. 
The canary sample consists of a sentence containing a \textit{secret code}. 
We then fine-tuned the model using this modified training set and evaluated its perplexity on the validation set. 
After fine-tuning, we provided a prompt (specifically, a partial prefix of the complete canary sentence) to generate 1,000 unique candidate outputs.
We will detail the generation method in the following subsection.
Next, we computed the cross-entropy loss for each candidate relative to the true canary sample, ranked these losses to obtain a rank, and finally used this rank to calculate the exposure detailed in \textbf{Definition 1.} 
A higher exposure indicates that the model has a stronger memorization of the canary, implying a higher privacy risk.
We employed two kind of attacks.
In the weak attack, the model is provided only with the prefix that precedes the \textit{secret code} and is expected to generate the complete \textit{secret code}. 
In the strong attack, the model is challenged to output only the final character of the \textit{secret code}.
\subsubsection{Definition 1. Exposure}
\begin{equation}
\text{\textbf{exposure}}_f(s) = \log_2 \lvert C \rvert - \log_2 \bigl(\text{\textbf{rank}}_f(s)\bigr)
\end{equation}
We follow the definition \cite{48112} in our study. The candidate space $C$ represents the number of candidates generated during the generation rather than every possible character combination of the same length as the secret string \textit{s}.
\subsubsection{Candidates Generation.}
Given that the target canary and candidate outputs share a common prefix,
we use stochastic decoding techniques to generate a diverse set of candidate texts. 
Specifically, we employ sampling strategies such as temperature scaling,
top-k, and nucleus (top-p) sampling to produce multiple outputs from the fine-tuned model. 
We apply truncation to ensure that all candidate outputs maintain the same length.
This generation process is iterated until a predefined number of unique candidate texts is obtained, ensuring a comprehensive representation of
the model’s output distribution under similar input conditions.
\subsubsection{Design Motivation.}
In data extraction research, exposure is frequently used as a risk metric. In Carlini et al. \cite{48112} 's original study, the candidate space was defined as all possible character combinations. However, when the target string is long, this approach results in an enormous candidate space that is impractical to compute. 
An alternative \cite{marchyok2025evaluating} involves manually selecting similar targets from the dataset, such as similar English names, to form the candidate space. 
Yet, this method requires extensive data processing and remains impractical for large datasets. To address these challenges, we designed a method in which the model itself generates candidate outputs. 
It is important to emphasize that, strictly speaking, our approach is not a true attack but rather an evaluation technique to assess the model's risk of memorizing sensitive information. 
By simulating an adversary using a prompt to extract data, we can measure the model's exposure and thus its privacy risk. Our experiments have demonstrated that this method is effective.
\subsection{Membership Inference Attack}
For the Membership Inference Attack, we adopted SPV-MIA. 
We implemented the attack using the open-source SPV-MIA code. To ensure compatibility with DP training and PEFT, we integrated fastDP and Hugging Face's PEFT library into the original code, enabling it to support PEFT-based DP training as well as perform inference and data generation.
\subsubsection{Adversary's Capability.}
The adversary in SPV-MIA operates in a black-box environment, meaning they can only submit crafted prompts and observe the model’s outputs. 
Additionally, it is assumed that the adversary may have partial information about the training data distribution.
\subsubsection{Threat Model.}
Instead of relying on pre-existing data, the adversary uses the target language model to generate candidate reference texts that approximate the training data’s distribution.
These self-generated texts are then used to fine-tune a reference model. 
The core of SPV-MIA is a probabilistic variation metric that quantifies how much the target model memorizes a specific record. 
This metric is defined as the expectation of the second-order directional derivative of the model's probability function:
\begin{equation}
\tilde{p}_\theta(x) := \mathbb{E}_z\left[ z^\top H_p(x) z \right],
\end{equation}
where \( H_p(x) \) is the Hessian of the probability function \( p_\theta(x) \).
\begin{equation}
A_{\text{our}}(x, \theta, \hat{\theta}) = 1\left[ \tilde{p}_\theta(x) - \tilde{p}_{\hat{\theta}}(x) \geq \tau \right],
\end{equation}
where \( \hat{\theta} \) is the self-prompt reference model, and \( \tau \) is the threshold for membership inference.

The difference serves as a robust membership signal that does not solely depend on overfitting.

\section{Experiments}
\subsection{Key Findings Takeaway}
The following key findings summarize the main trends observed in our experiments and provide guidance on the trade-offs between model utility and privacy protection:
\begin{itemize}
  \item[(1)] Differential privacy reduces model utility, with lower privacy budgets causing greater degradation. Full fine-tuning and LoRA exhibit strong robustness, whereas prefix-tuning suffers significantly, particularly in larger models.
  \item[(2)] Without DP, full fine-tuning and LoRA are prone to extreme memorization, resulting in very high exposure and MIA risk, while prefix-tuning and P-tuning naturally offer better privacy protection.
  \item[(3)] With DP training, both exposure and MIA risk are significantly reduced across all methods, with full fine-tuning and LoRA benefiting most; however, further lowering the privacy budget has only marginal effects on MIA AUC.
  \item[(4)] In terms of privacy-utility trade-off, full fine-tuning achieves the best overall balance. Among the PEFT approaches, LoRA excels in preserving utility and P-tuning offers superior privacy protection, whereas prefix-tuning is not recommended due to its severe utility loss under DP.
\end{itemize}

\subsection{Experimental Setup}
\subsubsection{Environment.}
The experiments were conducted on a Linux server running Ubuntu 22.04.5 LTS. 
Our environment used Python 3.10 with PyTorch 1.13.0 and CUDA 11.6, alongside key libraries such as Transformers and fastDP. 
The server featured dual NVIDIA A6000 GPUs (totaling 96GB of VRAM), an AMD EPYC 7313P CPU, and 503GB of system memory, providing a robust platform for training and evaluating our models under differential privacy and various fine-tuning methods.

\subsubsection{Datasets.}
In our study, we employed two datasets for fine-tuning the language model: \textbf{Wikitext-2-v1}~\cite{merity2016pointer} and \textbf{AG News}~\cite{Zhang2015CharacterlevelCN}. 
\textbf{Wikitext-2-v1} is a high-quality corpus derived from English Wikipedia articles. 
It contains 36718 samples for training and 3760 samples for validation.
% It has been carefully curated and preprocessed for language modeling tasks, providing a rich collection of well-formed and diverse text. 
\textbf{AG News} is a widely recognized dataset used primarily for text classification tasks.
It contains 120000 samples for training and is known for its concise yet diverse content.

For the data extraction attack, we utilize the entire Wikitext-2-v1. 
The canary sample is \textit{The secret code is hzdh0831.}
The method for constructing the training set is detailed in methodology.
In the Membership Inference Attack, we extract 10,000 samples from the AG News' train-set to form the training set, which represents the member set in the context of the attack. 
Additionally, we extract 1,000 samples from AG News' test-set to serve as the evaluation set, which simultaneously functions as the non-member set.
\subsubsection{Models.}
We employ two autoregressive language models: GPT-2 and GPT-2 XL~\cite{radford2019language}. 
GPT-2 is an autoregressive model based on the transformer architecture, pre-trained on a diverse corpus of web text. 
It is well-known for its ability to generate coherent and contextually relevant text, making it a popular choice for language modeling tasks. 
GPT-2 XL, the largest variant in the GPT-2 family, contains a significantly higher number of parameters, which allows it to capture more intricate language patterns and deliver enhanced performance on complex tasks.
GPT-2 has 124M parameters while GPT-2 XL has 1.5B parameters.

For the fine-tuning methods, we adopt full fine-tuning as our baseline. 
Additionally, we explore three popular parameter-efficient fine-tuning (PEFT) approaches: prefix-tuning, LoRA, and P-tuning, which are detailed in Section 3. 
For PEFT of GPT-2, we use 8 as the rank of LoRA, 30 as the length of virtual tokens in Prefix-tuning and P-tuning, 128 as the size of hidden encoder in P-tuning.
These methods are implemented using the peft~\cite{peft} module available in the Transformers~\cite{wolf-etal-2020-transformers} library, enabling us to compare their performance in terms of utility and privacy risks.
In two attack experiments, we fine-tuned for 10 epochs for the target model and used early-stopping to prevent overfitting.

\subsubsection{Differential Privacy Mechanism.}
We employ DP-Adam to provide sample-level differential privacy protection for the fine-tuning dataset. 
Specifically, we utilize fastDP~\cite{bu2023differentially,bu2022differentially,bu2023zero}, a library that enables differentially private optimization of PyTorch models. 
FastDP implements gradient clipping based on Book-Keeping approach to achieve differential privacy. 
It can integrate seamlessly with our chosen PEFT methods and language models.

\subsubsection{Metrics.}
For model utility analysis, we use perplexity as the evaluation metric in both experiments. 
In the prompt attack experiment, we measure the risk of privacy leakage using the exposure metric, which is detailed in Section 3. 
For the membership inference attack (MIA) experiment, we employ SPV-MIA as the attack method and assess its effectiveness using the AUC (Area Under the Curve) and ARC (Attack Success Rate) metrics.

\subsection{Impact of Differential Privacy on Model Utility}

\begin{figure}[t]
\centering
\includegraphics[width=\textwidth]{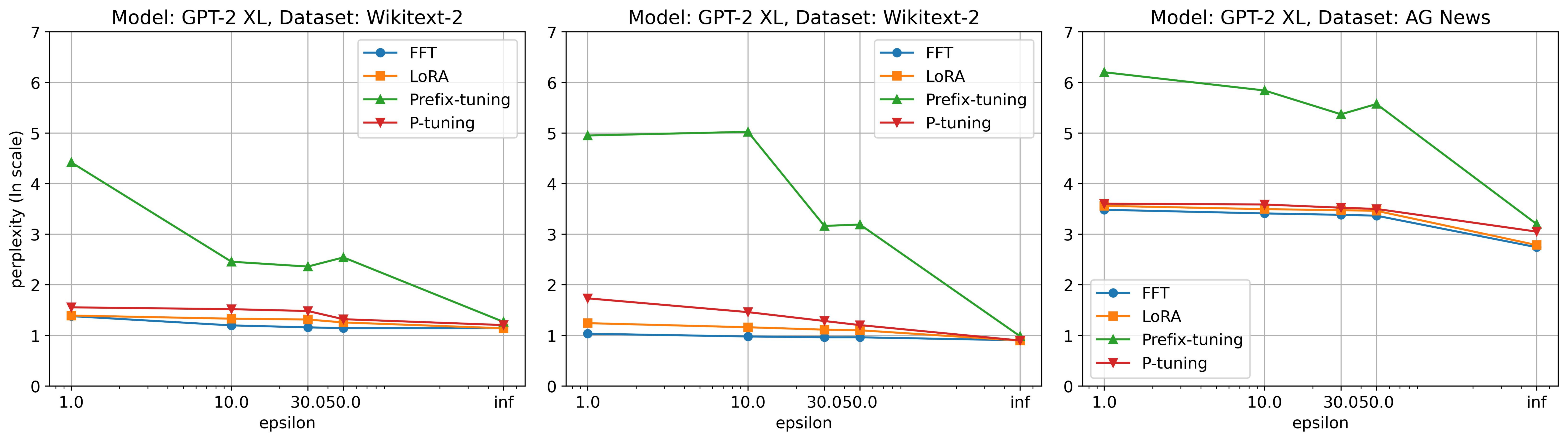}
\caption{\textbf{Perplexity as a function of the privacy budget $\epsilon$ for various fine-tuning methods.} The x-axis represents the privacy budget, while the y-axis shows the model perplexity in \textit{ln} scale (with lower values indicating higher utility). Trends in the figure illustrate the impact of different DP budgets on model utility across various fine-tuning methods.} \label{fig1}
\end{figure}

We plotted the perplexity of fine-tuned models against different privacy budgets across two settings in Fig. \ref{fig1}. Without the application of differential privacy, all three PEFT methods exhibit perplexity levels comparable to full fine-tuning, with LoRA even matching the full fine-tuning performance. This indicates that, in the absence of DP, the selected PEFT approaches can maintain high utility, with LoRA showing the best performance.

However, once noise is introduced via differential privacy, the utility of the models degrades across all fine-tuning methods. As the privacy budget decreases (i.e., as more noise is injected), the perplexity of the models increases. 
We fine-tuned both GPT-2 and GPT-2 XL on Wikitext-v2. The results reveal that, with the exception of prefix-tuning, all fine-tuning methods yield lower perplexity on DP-trained GPT-2 XL compared to GPT-2, indicating enhanced utility for larger models. However, prefix-tuning exhibits the opposite behavior. 
GPT-2 XL fine-tuned with prefix-tuning shows higher perplexity, suggesting that differential privacy severely impacts its utility.
The anomalous behavior of prefix-tuning on larger models warrants further investigation and validation.
Notably, the extent of the perplexity's increase varies among the methods; prefix-tuning shows the most pronounced deterioration in utility, while FFT and LoRA continue to maintain relatively good performance.
A possible explanation for these observations is that methods like FFT and LoRA distribute the DP-induced noise across a larger set of parameters or incorporate it into low-rank updates, which mitigates its adverse effects. In contrast, prefix-tuning relies heavily on a small set of learned prompt representations to condition the model's output. Consequently, the noise has a disproportionate impact on these few parameters, leading to a significant decline in their effectiveness and, therefore, a marked drop in utility.

% \fbox{\parbox{\textwidth}{

% }}
\begin{remark}
Differential privacy reduces model utility, with lower privacy budgets causing greater degradation. LoRA and full fine-tuning are relatively robust, while prefix-tuning suffers severe utility loss. Notably, although larger models typically yield higher utility at the same privacy budget, prefix-tuning exhibits a counterintuitive drop.
\end{remark}

\subsection{Impact of Differential Privacy on Empirical Privacy Risk}
To evaluate the impact of differential privacy on models' empirical privacy risk. We performed two attacks on GPT-2 obtained by different fine-tuning methods which is detailed in Section \ref{section3}.

\subsubsection{Data Extraction.}
\begin{figure}
\centering
\includegraphics[width=0.7\textwidth]{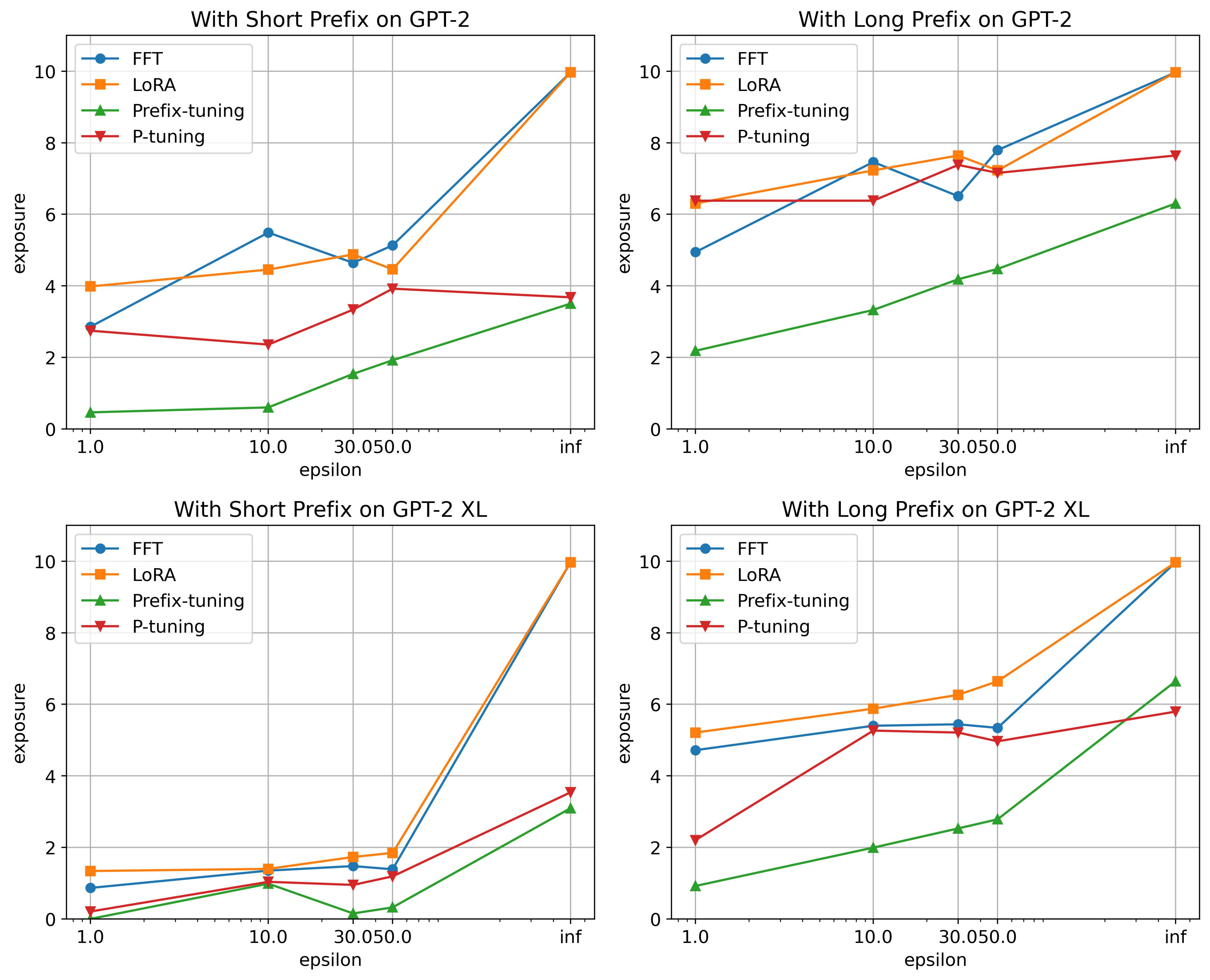}
\caption{\textbf{Exposure as a function of the privacy budget $\epsilon$ for different fine-tuning methods.} The four subplots display results for two model sizes and two attack scenarios. The top two subplots correspond to GPT-2, while the bottom two correspond to GPT-2 XL. The left subplots represent the weak attack (Short Prefix), and the right subplots represent the strong attack (Long Prefix). These trends illustrate how DP training affects model exposure across varying privacy budgets and attack strengths.} \label{fig2}
\end{figure}

We plotted the exposure metric against the privacy budget $\epsilon$ for various fine-tuning methods in Fig. \ref{fig2}.

Without applying any DP mechanism, both FFT (full fine-tuning) and LoRA exhibit extremely high exposure, reaching the maximum possible level. 
This indicates that these models can directly output the secret code embedded in the canary, thus posing a significant privacy risk. 
An adversary can easily extract sensitive information by simply providing an appropriate prompt. 
In contrast, prefix-tuning and P-tuning inherently offer a degree of privacy protection, as their exposure is much lower than FFT and LoRA with no DP mechanism 
We hypothesize that it is related to the location of the fine-tuned parameters within the model.
Methods like FFT and LoRA modify parameters that are distributed across key layers (such as the self-attention and feed-forward networks) which are directly involved in generating outputs and encoding detailed information from the training data. 
This broader and deeper integration allows these methods to capture and retain more specific patterns and even sensitive details. 
In contrast, techniques like prefix-tuning and P-tuning adjust only a small set of additional parameters—typically in the form of prompt embeddings or similar auxiliary tokens—that are less tightly coupled with the core model representations. 
This more limited and peripheral update results in a lower capacity for memorization.

After introducing DP noise, the results show distinct behaviors across the fine-tuning methods. For FFT and LoRA that tend to develop strong memorization of the training data, the application of DP leads to a significant reduction in exposure even at a high privacy budget (e.g., $\epsilon$=50). This indicates that DP is highly effective in mitigating the memorization—and thus the privacy risk—of sensitive data in these models. 
Conversely, for prefix-tuning and P-tuning, although the introduction of DP noise does lower exposure, the reduction is less pronounced. 
We speculate that this may be because these two methods, particularly in smaller models like GPT-2, do not form strong memorization in the first place, leaving less room for DP to further reduce exposure.
Furthermore, as the privacy budget decreases, there is a general trend of declining exposure, which suggests that a stricter privacy budget does indeed enhance privacy protection. 
However, compared to the initial drop in exposure observed when DP was first applied, the subsequent decreases are not as significant. 

Another interesting finding is that, when mitigating prompt attacks, differential privacy proved more effective on GPT-2 XL than on GPT-2, as shown by a more pronounced decrease in exposure. 
One possible explanation is that larger models have more parameters, allowing them to better absorb and distribute the noise introduced by DP. 
This dilution of noise effects on individual parameters reduces the model's tendency to memorize specific training data, thereby enhancing privacy protection under the same privacy budget. 

\begin{remark}
\label{remark:privacy budget}
    Without differential privacy, full fine-tuning and LoRA result in very high exposure, which contradicts previous studies that reported LoRA as having the lowest exposure. 
    In contrast, prefix-tuning and P-tuning naturally provide better privacy protection. 
    When differential privacy is applied, exposure decreases significantly, and lower privacy budgets lead to even lower exposure. 
    Moreover, more complete prompts tend to increase privacy risk. Larger models exhibit lower exposure than smaller ones under the same privacy budget.
\end{remark}

% It should be noted that, in our experiments on GPT-2 XL, we did not use the prefix-tuning model as an attack target. 
% Under DP, the GPT-2 XL model with prefix-tuning exhibited excessively high perplexity, leading to very low-quality generated text that was practically unusable. 
% Consequently, the exposure metric in this scenario does not hold meaningful value.
% In our analysis of the impact of the privacy budget on model utility, we observed that prefix-tuning is particularly sensitive to privacy noise. 
% This sensitivity is especially pronounced in larger models like GPT-2 XL, where the noise significantly degrades overall utility.
\subsubsection{Membership Inference Attack.}

\begin{figure}[t]
\centering
\includegraphics[width=0.5\textwidth]{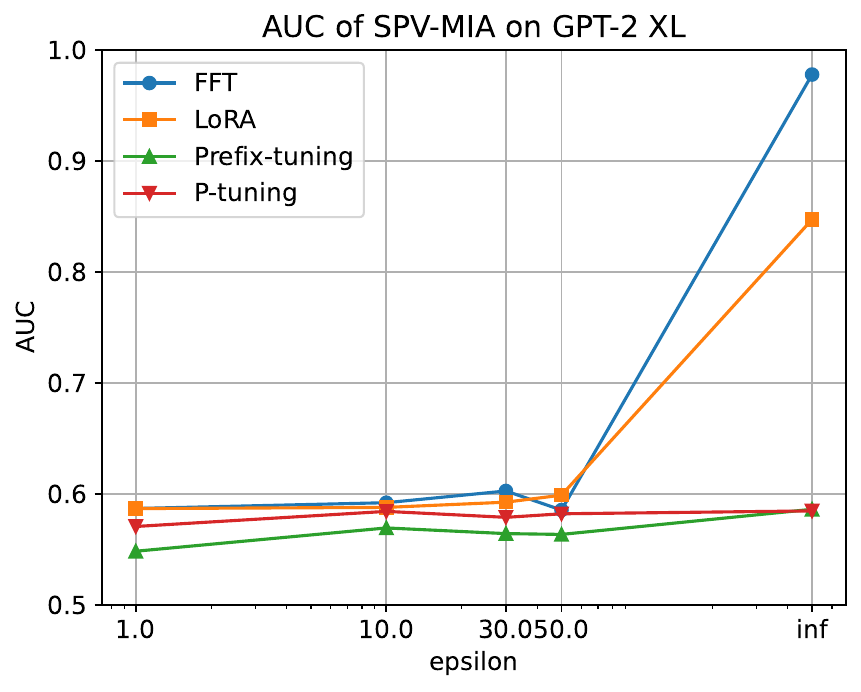}
\caption{\textbf{MIA attack performance as a function of privacy budget $\epsilon$ on GPT-2 XL.} This plot illustrates how the effectiveness of membership inference attacks (as measured by AUC) varies with different privacy budgets. Each point represents the MIA performance under a specific $\epsilon$ value.} \label{fig3}
\end{figure}

\begin{figure}[t]
\centering
\includegraphics[width=0.9\textwidth]{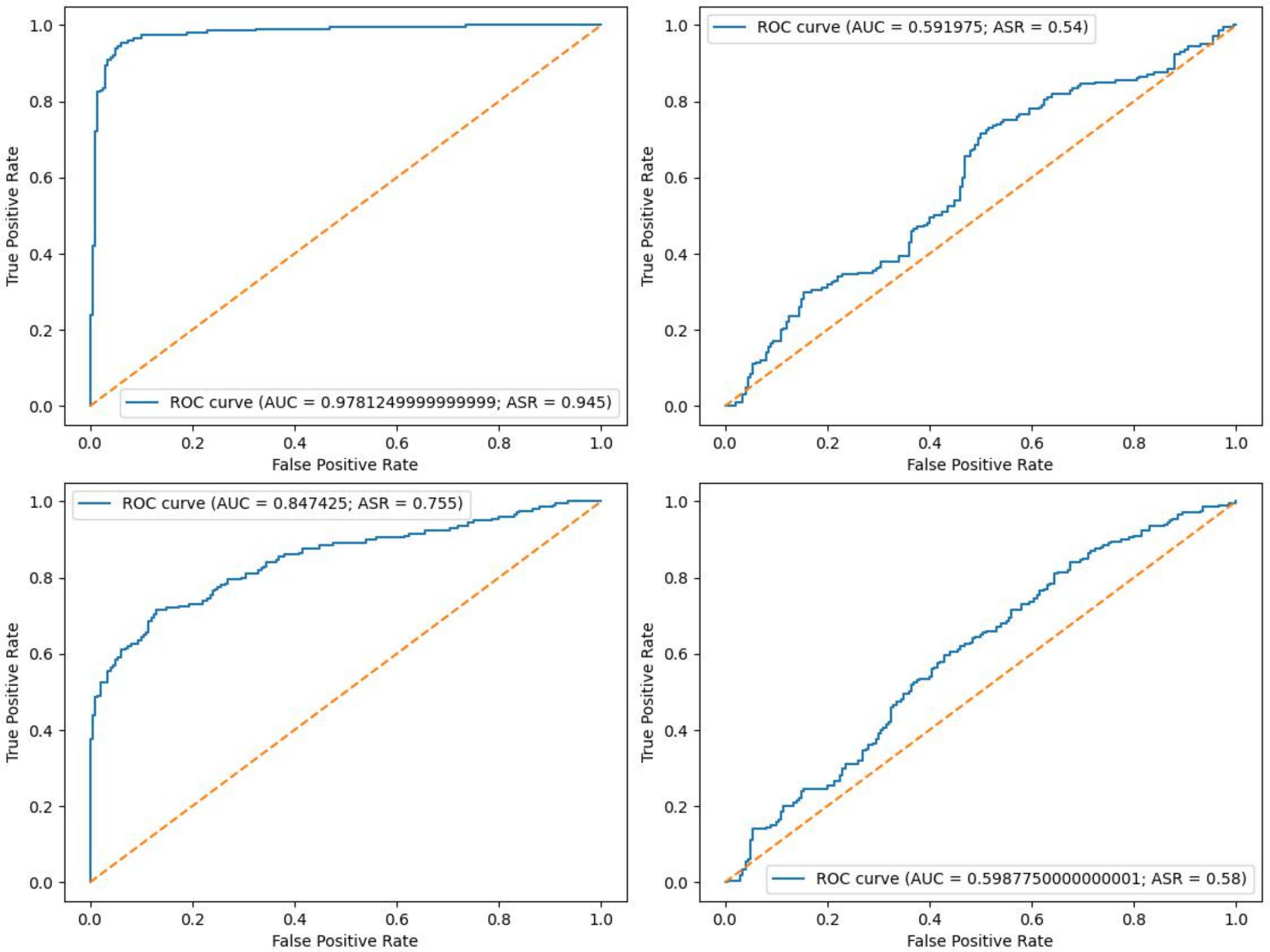}
\caption{\textbf{ROC Curves for Membership Inference Attacks (MIA) on Models with and without Differential Privacy (DP) Training.} The top two subplots show the ROC curves for full fine-tuning (FFT) before DP training and after DP training (with $\epsilon$=10), while the bottom two subplots depict the corresponding results for LoRA. The ASR (Attack Success Rate) is indicated in the plots. This figure demonstrates that DP training significantly mitigates the effectiveness of MIA attacks, as reflected in the substantial degradation of the ROC curves when DP is applied.} \label{fig4}

\end{figure}
In our MIA experiments, we observed some intriguing phenomena and we plot the results in Fig. \ref{fig3}. Without differential privacy (DP), both FFT and LoRA were extremely vulnerable to MIA, achieving AUC values exceeding 80\% and FFT reaching as high as 97.8\%, which clearly indicates a very high privacy risk. 
We attribute this vulnerability to the strong memorization capabilities inherent in FFT and LoRA, leading these methods to effectively capture and reproduce sensitive information from the training data. 
In contrast, prefix-tuning and P-tuning exhibited substantially lower AUC values (around 60\%) in the absence of DP, consistent with our findings from the prompt attack experiments. 
To eliminate the influence of model architecture, we fine-tuned GPT-J \cite{gpt-j} on the same dataset without differential privacy. 
The results show that GPT-J fine-tuned with P-tuning and prefix-tuning still maintains a low AUC of around 0.6 against SPV-MIA.
These results suggest that prefix-tuning and P-tuning naturally offer a degree of privacy protection by limiting the memorization of individual training samples, which is consistent with our findings in the data extraction.

Upon introducing a differential privacy mechanism, all four fine-tuning methods experience a substantial reduction in AUC. In particular, FFT and LoRA see dramatic decreases, with all methods converging to approximately 58\% AUC. We plotted the ROC curves of GPT-2 XL under SPV-MIA before and after DP training in Fig. \ref{fig4}. It can be seen that DP training significantly reduces the attack threat on LoRA and full fine-tuning.

However, further lowering the privacy budget beyond this point does not yield a more pronounced decline in AUC. 
One potential explanation for this plateau is that the initial application of differential privacy introduces enough noise to disrupt the memorization signals exploited by the attacks, effectively reducing vulnerability across the board. Once these signals are sufficiently obfuscated, additional noise, which is achieved by further lowering the privacy budget, offers diminishing returns in terms of additional privacy gains. 
In fact, even with a privacy budget of $\epsilon$=1, differential privacy guarantees that the success rate of MIA remains below 73\%, which is still higher than the AUC values observed in current MIAs.
This may imply that current MIA methods are still insufficient in their ability to effectively attack models trained with differential privacy.

\begin{remark}
Without differential privacy, full fine-tuning and LoRA exhibit extremely high MIA risk, whereas prefix-tuning and P-tuning inherently offer some privacy protection, consistent with our previous findings. 
Once DP training is introduced, the MIA risk decreases for all fine-tuning methods, with reductions being particularly significant for full fine-tuning and LoRA. 
Notably, even at very high privacy budgets, DP still provides substantial protection against MIA, though further changes in the privacy budget have little impact on the AUC.
\end{remark}

\subsection{Trade-off between Utility and Privacy}

\begin{figure} [t]
\centering
\includegraphics[width=0.9\textwidth]{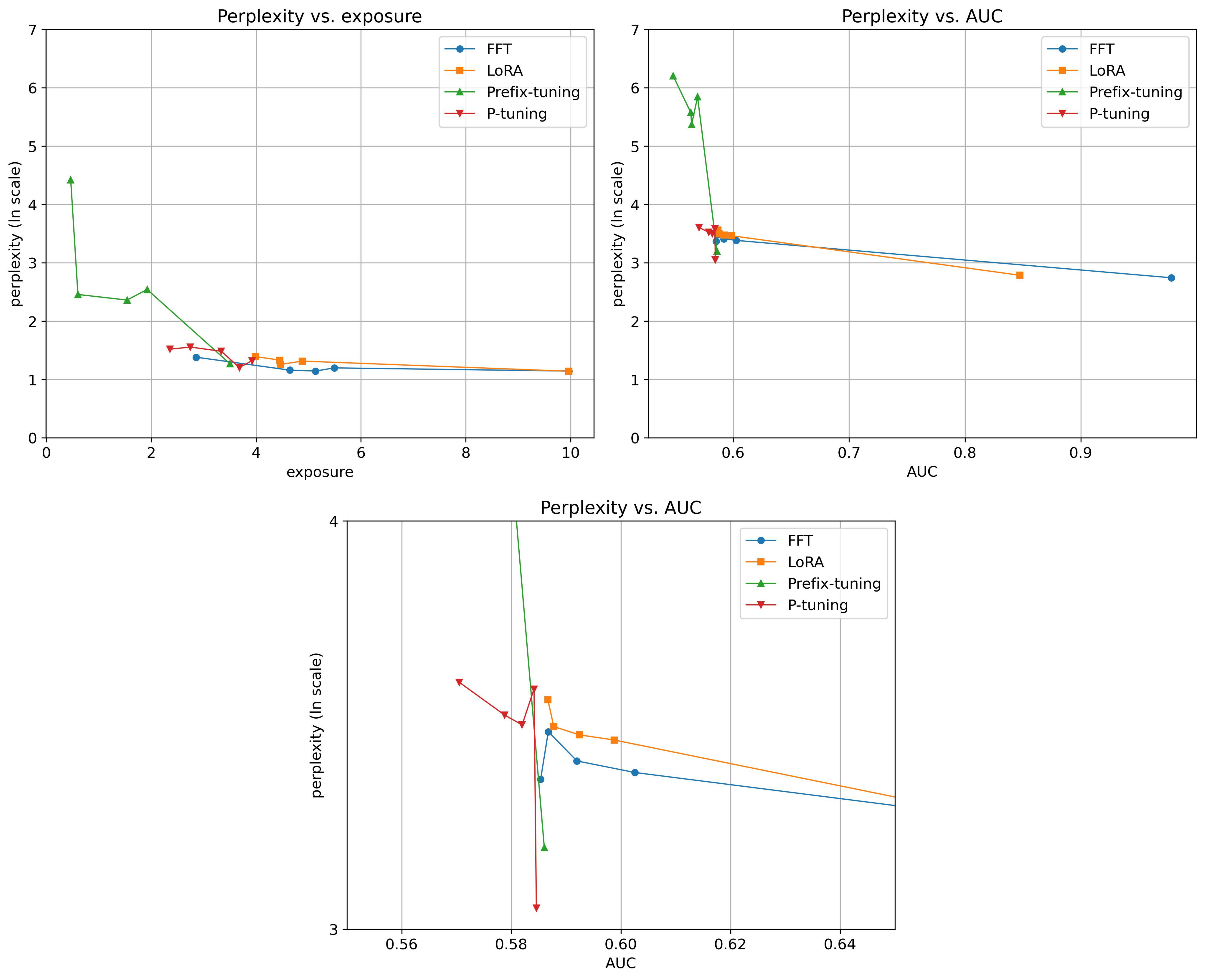}
\caption{\textbf{Trade-off between model privacy risk and utility.} The left panel plots exposure versus perplexity, while the right panel shows AUC versus perplexity, with perplexity on the vertical axis in both plots. In each curve, the rightmost point corresponds to the model trained without differential privacy. We have enlarged part of the right panel to better observe the relationship between the curves.} \label{fig5}
\end{figure}

Achieving an optimal trade-off between utility and privacy remains a critical objective. To evaluate how different fine-tuning methods balance utility and privacy protection under differential privacy (DP), we plotted the relationship between exposure and perplexity as well as AUC and perplexity for each fine-tuning method, as shown in Fig. \ref{fig5}. 
In these plots, every point on a curve corresponds to the model's performance under a specific privacy budget. 
We interpret a better privacy-utility trade-off as a situation where a model maintains low perplexity while also keeping exposure and AUC low; therefore, curves that are generally closer to the lower-left corner indicate superior trade-offs. 
It is important to note that the rightmost point on each curve represents $\epsilon$ = $\infty$ (i.e., no DP), and when considering trade-offs under differential privacy, this point can be disregarded.

Based on the plots, under differential privacy the curves for full fine-tuning are consistently closer to the lower-left corner in both graphs. This indicates that full fine-tuning achieves the best privacy-utility trade-off under DP. We attribute this to the large number of trainable parameters in full fine-tuning, which enables the model to better absorb and mitigate the impact of DP-induced noise while still protecting privacy.

Among the PEFT methods, both LoRA and P-tuning demonstrate a favorable trade-off. Specifically, the curve for P-tuning is positioned further left, suggesting a stronger advantage in privacy protection, whereas the curve for LoRA is shifted slightly to the right, reflecting better utility preservation. In contrast, prefix-tuning under DP leads to excessively high perplexity, severely compromising model utility. Although it offers relatively strong privacy protection, the excessive loss in utility—likely due to the model's diminished capacity to memorize—renders prefix-tuning unsuitable for DP training in its current form. This suggests a need for developing DP training methods tailored specifically for prefix-tuning.

In summary, full fine-tuning appears to be the best option under DP due to its high privacy-utility trade-off. However, given the substantial computational resources and time required for full fine-tuning, it is often impractical with large models. Among the PEFT approaches, LoRA is recommended when model utility is the primary concern, whereas P-tuning is preferable when stronger privacy protection is desired. We do not recommend using prefix-tuning under DP training because of its severe impact on utility. 
It is important to note that prefix-tuning and P-tuning naturally offer a certain degree of privacy protection even without DP. 
Therefore, if reducing privacy risk is the primary concern and DP training is not feasible, these methods are attractive alternatives.

\begin{remark}
Under differential privacy, full fine-tuning achieves the best privacy-utility trade-off, as it maintains low perplexity and low exposure/AUC. 
Among the PEFT methods, LoRA and P-tuning offer promising results. 
P-tuning excels in privacy protection, while LoRA better preserves utility. 
In contrast, prefix-tuning suffers from excessively high perplexity under DP, severely impairing utility, and is therefore not recommended. 
These findings suggest that while full fine-tuning is optimal, practical constraints make PEFT methods attractive; among them, LoRA is preferable when utility is paramount, and P-tuning when privacy is the main concern.
\end{remark}

\section{Future Work}
While our study offers valuable insights into the impact of privacy budgets on model utility and protection across fine-tuning methods, it is limited in several ways.
Our experiments focused solely on the GPT-2 family, which may not capture the full range of behaviors exhibited by other architectures or larger language models.
Additionally, we limited our analysis to standard differentially private optimization algorithms, even though recent advances have introduced specialized DP methods for PEFT, such as RAFT~\cite{li2023privacyprompttuning}, which might offer improved trade-offs between utility and privacy.
Our exploration of PEFT parameter configurations was also restricted to a few settings, leaving open questions about how different configurations might affect model performance and privacy risk. Furthermore, our evaluation was based on a narrow set of datasets, which may not fully represent the diversity of real-world data distributions and their associated privacy challenges. 
Future work should address these limitations by incorporating a broader range of models, exploring diverse DP methods (especially those tailored for PEFT), systematically varying parameter settings, and utilizing more varied datasets to enhance the generalizability and robustness of the findings.

\section{Conclusion}
In this work, we investigated the impact of differential privacy on the utility and empirical privacy risk of large language models fine-tuned with various methods, including both full fine-tuning and parameter-efficient fine-tuning methods.
Our experiments demonstrate that while differential privacy effectively reduces model exposure and membership inference risks, it also degrades utility, with the extent of degradation varying significantly across different fine-tuning methods.
Moreover, our analysis of the privacy-utility trade-off reveals that, under DP, full fine-tuning achieves the best overall balance; however, practical considerations such as resource constraints make PEFT methods, particularly LoRA and P-tuning, attractive alternatives depending on whether the emphasis is on preserving utility or ensuring privacy.
Overall, our findings provide valuable guidance for selecting fine-tuning strategies under DP constraints and set the stage for future research aimed at optimizing the trade-off between model performance and privacy protection in large-scale language models.

\vspace{5em}
%
% ---- Bibliography ----
%
% BibTeX users should specify bibliography style 'splncs04'.
% References will then be sorted and formatted in the correct style.
%
\bibliographystyle{splncs04}
\bibliography{Reference}
%
% \begin{thebibliography}{8}
% \bibitem{ref_article1}
% Author, F.: Article title. Journal \textbf{2}(5), 99--110 (2016)

% \bibitem{ref_lncs1}
% Author, F., Author, S.: Title of a proceedings paper. In: Editor,
% F., Editor, S. (eds.) CONFERENCE 2016, LNCS, vol. 9999, pp. 1--13.
% Springer, Heidelberg (2016). \doi{10.10007/1234567890}

% \bibitem{ref_book1}
% Author, F., Author, S., Author, T.: Book title. 2nd edn. Publisher,
% Location (1999)

% \bibitem{ref_proc1}
% Author, A.-B.: Contribution title. In: 9th International Proceedings
% on Proceedings, pp. 1--2. Publisher, Location (2010)

% \bibitem{ref_url1}
% LNCS Homepage, \url{http://www.springer.com/lncs}, last accessed 2023/10/25
% \end{thebibliography}
\end{document}